\documentclass[12pt]{iopart}
\usepackage{iopams}
\usepackage{tikz}
\usetikzlibrary{calc}
\usepackage{tikz}
\usepackage[latin1]{inputenc}
\usetikzlibrary{calc,arrows,positioning}
\usetikzlibrary{trees}
\usetikzlibrary{decorations.pathmorphing}
\usetikzlibrary{decorations.markings}
\newcommand{\sca}[2]{\langle #1,\, #2\rangle}
\newcommand{\ben}{\begin{equation}}
\newcommand{\bens}{\begin{equation*}}
\newcommand{\bal}{\begin{align}}
\newcommand{\een}{\end{equation}}
\newcommand{\eens}{\end{equation*}}
\newcommand{\eal}{\end{align}}
\newcommand{\bea}{\begin{eqnarray}}
\newcommand{\eea}{\end{eqnarray}}
\newcommand{\beas}{\begin{eqnarray*}}
\newcommand{\eeas}{\end{eqnarray*}}
\newcommand{\nn}{\nonumber\\ }

\newcommand{\q}{{\qquad}}
\newcommand{\qq}{\qquad\qquad}
\newcommand{\QQ}{\qquad\qquad\qquad\qquad}

\newcommand{\hfb}{{\hfill\break}}
\newcommand{\sub}[1]{_{\stackrel{}{#1}}}

\input amssym.def
\def\Z{{\mathbb Z}} \def\R{{\mathbb R}}  \def\N{{\mathbb N}}
   \def\P{{\mathbb P}}
\begin{document}
\voffset=-0.5cm

\title[]{Discretized Weyl-orbit functions:\\ modified multiplication and Galois symmetry}

\author{J Hrivn\'ak}

\address{Department of Physics, Faculty of Nuclear Sciences and Physical Engineering, Czech Technical University in
Prague, B\v rehov\' a 7, 115 19 Prague 1, Czech Republic\ }
\ead{jiri.hrivnak@fjfi.cvut.cz}

\author{M A Walton}

\address{Department of Physics and Astronomy, University of Lethbridge, Lethbridge, Alberta, T1K 3M4, Canada\ }
\ead{walton@uleth.ca}

\begin{abstract}
We note a remarkable similarity between the discretized Weyl-orbit functions and affine modular data associated with Wess-Zumino-Novikov-Witten (WZNW) conformal field theories. Known properties of the modular data are exploited here to uncover analogous results for the discretized orbit functions. We show that the product of orbit functions is modified in analogy with the truncation of tensor products known as affine fusion, governing the interactions in WZNW models. A Galois symmetry, like that of affine modular data,  is also described for the discretized orbit functions.
\end{abstract}

%\pacs{02.20.-a, 02.20.Qs, 02.20.Sv}
\date{\today}
%\maketitle

\section{Introduction}

The systematic study of Weyl-orbit sums as special functions was launched with the general work of  \cite{PstartI,PstartII,{KPc},KPs,{KPe}}. The simplest examples of Weyl-orbit functions (or just orbit functions, for short) are the $C$-function
\ben \Phi_\lambda(a) = \sum_{w\in W}\, e^{2\pi i\langle w\lambda, a\rangle}\ , \label{CfnIntro}
\een  and the $S$-function
\ben  \varphi_\lambda(a) =\sum_{w\in W}\, (\det w)\, e^{2\pi i\langle w\lambda, a\rangle}\ . \label{SfnIntro}\een Here $W$ is the Weyl group of the simple Lie algebra, $\langle\cdot,\cdot\rangle$ is the inner product in its weight space, and $a$ denotes a weight in a continuous region $F$ of weight space, to be specified below.

Discretized versions of these Weyl-orbit functions  were investigated in \cite{PstartII, disc}. They are the subject of this paper. One considers the weight $a$  in (\ref{SfnIntro}) not to belong to a continuum of possible values, but to a discrete fragment $F_M$ of a lattice inside the region $F$. Here $M$ is a positive integer, controlling the resolution of the discretized orbit function. Recently, other mathematical properties of the discretized Weyl-orbit functions and the discretization properties of the corresponding orthogonal polynomials have been studied intensively  (see \cite{xug2,xua2,munthe1,D},  e.g.).

Our key observation is that the discretized orbit functions are strikingly similar to important objects in conformal field theory. In particular, the Wess-Zumino-Novikov-Witten (WZNW) conformal field theories are each defined for a simple Lie algebra, as are the orbit functions. Furthermore, the WZNW model realizes an affine Kac-Moody algebra at a fixed  level, and so is determined by a simple Lie algebra and a positive integer (the level), just as are the discretized orbit functions.

The modular transformation matrices of any rational conformal field theory are pertinent both to the possible spectra of primary fields, and to their interactions \cite{DMS}. The torus partition function encodes the spectrum and must be modular invariant, and the fusion coefficients are determined by the modular $S$-matrix through the celebrated Verlinde formula. Kac and Peterson \cite{KaP} calculated the WZNW (or affine) modular $S$-matrix, finding
\ben  S_{\lambda,\mu} =  R_{M'}   \sum_{w\in W}\, (\det w)\, e^{-2\pi i \langle w\lambda, \mu\rangle/{M'}}\ \  \label{KPmatrixSIntro}\een  for its entries.  Here $R_{M'}$ is a constant fixed by the Lie algebra under consideration and the positive integer $M'$, while $\lambda$ and $\mu$ are shifted highest weights of integrable representations of the Lie algebra that are in 1-1 correspondence with representations of the affine Kac-Moody algebra at fixed level realized in the WZNW model.

The resemblance between the previous 2 equations is remarkable. Here we will exploit the similarity by using already-discovered properties of the affine  modular data  to uncover analogous attributes of the orbit functions.

More precisely, we will demonstrate that products of the discretized orbit functions decompose differently than the corresponding continuous functions. The modified multiplication is in perfect analogy with WZNW fusion, a modification (truncation) of the tensor product of representations of simple Lie algebras \cite{K, W}. In addition, the so-called Galois symmetry of the affine modular $S$-matrix \cite{CG} points to a similar Galois symmetry of the discretized orbit functions.

The next section  reviews the relevant Weyl-orbit functions and their discretized versions, while introducing necessary notation. Section 3 derives the modified multiplication of discretized orbit functions; Section 4 finds their Galois symmetries. Section 5 describes the motivation: the analogous (well-known) results for the affine modular data of WZNW conformal field theories. A short Conclusion follows.

\section{Weyl-orbit functions}

$G$ is a compact, complex, simply connected and connected simple Lie group, of rank $n$, and Lie algebra $g_n$. Its set of simple roots is denoted $\Pi = \{ \alpha_j \mid j\in I\}$, with $I:=\{1,\ldots, n\}$,  and they are normalized so that a long simple root $\alpha_{\rm long}$ obeys $\langle \alpha_{\rm long}, \alpha_{\rm long} \rangle=2$. The root lattice is $Q = \Z\Pi$. Each simple root $\alpha_j\in \Pi$ gives rise to a primitive reflection $r\sub{\alpha_j} = r_j$, in weight space, and the Coxeter-Dynkin diagram of $g_n$ encodes the defining relations of the Weyl group $W = \langle\, r_j\ \vert\ j\in I \,\rangle$. 

The fundamental weights of $g_n$ are denoted $\Omega = \{ \omega_j\ \vert\  j\in I \}$. They span the $n$-dimensional weight space $P_\R := \R\,\Omega \cong \R^n$. The cone of non-negative weights is $P_{+,\R}:= \R_{\ge 0} \,\Omega$, and $P_{++,\R}:=\R_{>0} \,\Omega$ is the cone of positive weights. The weight lattice is $P:=\Z \,\Omega$, and the cone of dominant (regular) integral weights is $P_+:= \N_0 \,\Omega$ ($P_{++}:= \N \,\Omega$).

The simple dual roots (co-roots) are $\alpha_j^\vee = 2\alpha_j/\langle\alpha_j, \alpha_j\rangle$.  They satisfy $\langle \alpha_i^\vee, \omega_j \rangle = \delta_{i,j}\ (i,j\in I)$.  The co-root lattice is $Q^\vee = \Z\Pi^\vee$, where $\Pi^\vee= \{ \alpha_j^\vee \mid j\in I\}$.

The objects of our study are the Weyl-orbit functions (\ref{CfnIntro}),  (\ref{SfnIntro}) where $\lambda, a\in P_\R$. $C$-functions (\ref{CfnIntro}) enjoy the $W$-invariance properties
\bens \Phi_{w\lambda}(a)\ =\ \Phi_{\lambda}(w a)\ =\ \Phi_\lambda(a)\ ,\ \forall w\in W.\     %\label{CW}
\eens We can therefore restrict the possible weights to $\lambda,\, a\in P_{+,\R}$, the fundamental region of $W$ in the weight space $P_\R= W P_{+,\R}$. $S$-functions are the antisymmetric Weyl-orbit sums (\ref{SfnIntro}). As a consequence of  their $W$-antisymmetry, \bens \varphi_{w\lambda}(a) = \varphi_{\lambda}(w a)= (\det w)\, \varphi_{\lambda}(a)\ ,\ \forall w\in W,\     \label{SW}\eens  $S$-functions vanish on the boundary of $P_{+,\R}$, and so only its interior $P_{++,\R}={\rm int}(P_{+,\R})$ is relevant.

The affine Weyl groups become relevant to the orbit functions when weights  $\lambda\in P_\R$ are restricted to be integral, i.e. $\lambda \in P$. Let $\xi$ indicate the highest root of $g_n$.  Augmented by $\alpha_0:=-\xi$, the simple roots $\Pi$  generate the extended Coxeter-Dynkin diagram of $g_n$.  Put $\Pi^{\rm aff} := \{ \alpha_j \mid j\in \hat I\}$. Here we have defined $\hat I := \{0,1,\ldots,n\} = I\cup \{0\}$. The extended Coxeter-Dynkin diagram of $g_n$ encodes the relations of the affine Weyl group \bens W^{\rm aff} = Q^\vee\, \rtimes\, W  =  \langle r_j \ \vert\  j\in\hat I \rangle\ . %\label{Waff}
\eens The 0-th simple reflection is given by
\bens r_0 a=r_\xi a + \frac{2\xi}{\sca{\xi}{\xi}}\ ,\qquad
r_{\xi}a=a-\frac{2\sca{a}{\xi} }{\sca{\xi}{\xi}}\xi\ .%\label{rzero}
\eens
For any $w^{\mathrm{aff}}\in W^{\mathrm{aff}}$, there exist a unique $w\in W$ and a unique shift $q^\vee\in Q^\vee $ such that $w^{\mathrm{aff}}a=wa+q^\vee$; the retraction homomorphism $\psi:{W}^{\mathrm{aff}}\rightarrow W$ is given by
\bens
\psi(w^{\mathrm{aff}}) =w.%\label{ret}
\eens
If $\xi = m_1\alpha_1+ \cdots + m_n\alpha_n$, the $m_j\in\N$ are known as {\it marks}. Putting $m_0=1$, we have a mark $m_i$  for every $i\in \hat I$. Taking $M$ a positive integer and defining the affine reflection $ r_{0,M}$ by
\bens r_{0,M}\, a = r_{\xi} a + M\frac{2\xi}{\sca{\xi}{\xi}}   ,%\label{rzeroM}
\eens
an augmented affine Weyl group $W^{\rm aff}_M$ is obtained as
\bens W^{\rm aff}_M = MQ^\vee\, \rtimes\, W \  =\ \langle\, r_{0,M}, r_1, \ldots, r_n \,\rangle\ .
%\label{WaffM}
\eens
For any $w^{\mathrm{aff}}\in W^{\mathrm{aff}}_M$, there exist a unique $w\in W$ and a unique shift $q^\vee\in Q^\vee $ such that $w^{\mathrm{aff}}a=wa+Mq^\vee$; the retraction homomorphism $\psi: {W}^{\mathrm{aff}}_M\rightarrow W$ is given by
\bens
\psi(w^{\mathrm{aff}}) =w.%\label{0retM}
\eens

If the arrows on the Dynkin diagram of $G$ are reversed, so that short (long) simple roots become long (short) ones, we find the Coxeter-Dynkin diagram of the dual Lie algebra $g^\vee_n$.  The reflections the dual roots define, $r^\vee_j = r_j\ (j\in I)$, generate the same Weyl group $W$, but a different affine Weyl group, called the dual affine Weyl group:
\bens \widehat W^{\rm aff} = Q\, \rtimes\, W  = \langle\, r_j^\vee\ \vert\ j\in\hat I \,\rangle = \langle\, r_0^\vee, r_1, \ldots, r_n \,\rangle\ . %\label{Whataff}
\eens
Here
\bens r_0^{\vee} a=r_{\eta} a + \frac{2\eta}{\sca{\eta}{\eta}}\ , \q r_{\eta}a=a-\frac{2\sca{a}{\eta} }{\sca{\eta}{\eta}}\eta\ ,%\label{rvzero}
\eens  where $\eta$ is the highest dual root.  The expansion $\eta =: -\alpha^\vee_0 = m_1^\vee\alpha^\vee_1+ \ldots + m_n^\vee\alpha^\vee_n$ defines the {\it dual marks} $m^\vee_j,\ j\in I$, and we put $m_0^\vee=1$.
Taking $M$ a positive integer and defining the dual affine reflection $ r^\vee_{0,M}$ by
\bens
r^\vee_{0,M}\, a = r_{\eta} a + M\frac{2\eta}{\sca{\eta}{\eta}}   ,%\label{rvzeroM}
\eens
an augmented dual affine Weyl group $\widehat W^{\rm aff}_M$ is obtained as
\bens \widehat W^{\rm aff}_M = MQ\, \rtimes\, W   =\langle\, r_{0,M}^\vee, r_1, \ldots, r_n \,\rangle\ . %\label{WhataffM}
\eens
For any $w^{\mathrm{aff}}\in \widehat W^{\mathrm{aff}}_M$, there exist a unique $w\in W$ and a unique shift $q\in Q $ such that $w^{\mathrm{aff}}a=wa+Mq$; the dual retraction homomorphism $\widehat\psi:\widehat {W}^{\mathrm{aff}}_M\rightarrow W$ is given by
\bens
\widehat\psi(w^{\mathrm{aff}}) =w.%\label{retM}
\eens

The set of dual fundamental weights $\Omega^\vee := \{ \omega^\vee_j\ \vert\ j\in I \}$ satisfy $\langle \omega^\vee_j, \alpha_k \rangle =\ \delta_{j,k}$. They span the $n$-dimensional weight space $P_\R := \R\,\Omega \cong \R^n$ of $g^\vee_n$. The cone of non-negative dual weights is $P^\vee_{+,\R}:= \R_{\ge 0} \,\Omega^\vee$, and $P^\vee_{++,\R}:=\R_{>0} \,\Omega^\vee$ is the cone of positive dual weights. The dual weight lattice is $P^\vee :=\Z \,\Omega^\vee$, and the cone of dominant (regular) integral dual weights is $P^\vee_+ := \N_0 \,\Omega^\vee$ ($P^\vee_{++}:= \N \,\Omega^\vee$).

Now, for $\lambda\in P$, the $W$-symmetry of the Weyl-orbit functions is extended to affine Weyl symmetry for all $w\in W^{\rm aff}$:
\beas  \Phi_{\lambda}(w a)= \Phi_\lambda(a),\q \varphi_{\lambda}(w a) = \det \psi (w) \cdot \varphi_{\lambda}(a).%\label{CSWaff}
\eeas For this discretized case, we can restrict the domain of the $C$-functions to the fundamental region $F$ of $W^{\rm aff}$, the convex hull of the points $\{0, \frac{\omega_1^\vee}{m_1},\ldots, \frac{w_n^\vee}{m_n} \}$: \bens  F= \Bigl\{ \sum_{j\in I} a_j\omega_j^\vee \Bigm| a_j\in \R_{\ge0}\ ,\ \sum_{j\in \hat I} a_jm_j = 1 \Bigr\} .  %\label{F}
\eens Another notation would be $F = P_{+,\R}^{\vee,\, 1}$.  Since the $S$-functions vanish on the boundary of $F$, if $\lambda\in P$, it is only the interior $\widetilde F := {\rm int} (F) = P_{++,\R}^{\vee,\, 1}$ that is relevant for them.

Consider data with support $F$, or $\widetilde F$.  The Weyl-orbit functions provide useful expansion bases for the Fourier analysis of functions on $F$, or $\widetilde F$. Now suppose the data is digitized,  so that we are interested in the values on a discrete grid $F_M$ in $F$, or $\widetilde F_M$ in $\widetilde F$. $M\in \N$ will determine the resolution $\sim 1/M$ of the digital data of interest. Put
\bens%\label{PpM}
P_+^{\vee,\, M} := \Big\{\, \sum_{j\in I} a_j \omega^\vee_j\Bigm| a_j\in \N_0\ \forall j\in \hat I\, ,\ \sum_{j\in \hat I} a_jm_j\, =\, M  \,\Big\}\ ,
\eens and
\bens%\label{PppM}
P_{++}^{\vee,\, M} := \Big\{\, \sum_{j\in I} a_j \omega^\vee_j\Bigm| a_j\in \N\ \forall j\in \hat I\, ,\ \sum_{j\in \hat I} a_jm_j\, =\, M  \,\Big\}\ .
\eens
The discrete grid taking the place of the fundamental region $F$ is
\bens
F_M\ :=\frac 1 M\ P^{\vee,\, M}_{+}=\frac 1 M \ P^{\vee}\cap  F . %\label{FM}
\eens The corresponding interior of the grid $F_M$ is \bens \widetilde F_M :=\frac 1 M \ P^{\vee,\, M}_{++}=\frac 1 M \ P^{\vee}\cap \widetilde F.%\label{tFM}
\eens
But now if we have $a\in F_M$, we enjoy the dual affine Weyl invariance for all $\hat w\in \widehat W^{\rm aff}_M$:
\beas  \Phi_{\hat w\lambda}(a)=\Phi_\lambda(a) ,\q \varphi_{\hat w\lambda}(a)\ =\ \det \widehat\psi(\hat w) \cdot\,\varphi_\lambda(a) .%\label{CShWaff}
\eeas The weights $\lambda\in P_+$ can then be  restricted further.

The fundamental region of $\widehat W^{\rm aff}$, or dual fundamental domain, is $F^\vee$, $P^\vee_\R = \widehat W^{\rm aff} F^\vee$. It is the convex hull of vertices $\{0, \frac{\omega_1}{m^\vee_1},\ldots, \frac{\omega_n}{m_n^\vee} \}$:
\bens  F^\vee = \Big\{ \sum_{j\in I} \lambda_j\omega_j \Bigm| \lambda_j\in \R_{\ge0}\, \forall j\in\hat I ,\ \sum_{j\in \hat I} \lambda_jm_j^\vee = 1 \Big\}\ ,  %\label{Fv}
\eens
 i.e., $F^\vee = P_\R^1$.

With discretized data, of resolution $\sim 1/M$, we expect the Fourier-dual weights $\lambda$ to be dilated by a factor of $M$. For $C$-functions $\Phi_\lambda(a)$, we can restrict to $\lambda\in \Lambda_M$:
\bens \Lambda_M := MF^\vee\, \cap P= P_+^M\  .%\label{LaM}
\eens Here \bens  P_+^M := \Big\{\, \sum_{j\in I} a_j \omega_j\Bigm| a_j\in \N_0\ \forall j\in \hat I\, ,\ \sum_{j\in \hat I} a_jm^\vee_j\, =\, M  \,\Big\}\ .%\label{PnovpM}
\eens The grid $\Lambda_M$ is a truncation of the cone $P_+ = \N_0 \,\Omega$ of dominant weights.  The boundary points of $\Lambda_M$ produce vanishing $S$-functions.
The labels of $S$-functions $\varphi_\lambda(a)$ are therefore restricted to the interior $\widetilde \Lambda_M$ of $\Lambda_M$:
\bens \widetilde\Lambda_M = {\rm int}(\Lambda_M) = P_{++}^M\ .%\label{tLM}
\eens

The affine Weyl symmetry of the orbit functions can be summarized as follows. With $a\in F_M$ and  $\lambda\in \Lambda_M$, we have
\bea \Phi_\lambda(w a) &= \Phi_\lambda(a),\ \ w\in W^{\rm aff}\ ;\nn
 \Phi_{\hat w\lambda}(a) &=\Phi_\lambda(a),\ \ \hat w\in {\widehat W}^{\rm aff}_M\ ;
\eea
and with $\tilde a\in \widetilde F_M$ and  $\tilde\lambda\in \widetilde\Lambda_M$
\bea
 \varphi_{\tilde\lambda}(w\tilde a) &= \det \psi(w)\cdot\, \varphi_{\tilde\lambda}(\tilde a) ,\ \ w\in W^{\rm aff}\ ;\nn
 \varphi_{\hat w\tilde\lambda}(\tilde a) &=\det \widehat \psi(\hat w)\cdot\varphi_{\tilde\lambda}(\tilde a) ,\ \ \hat w\in {\widehat W}^{\rm aff}_M.\label{CSWaffhWaff}\eea

\begin{figure}
\resizebox{15.5cm}{!}{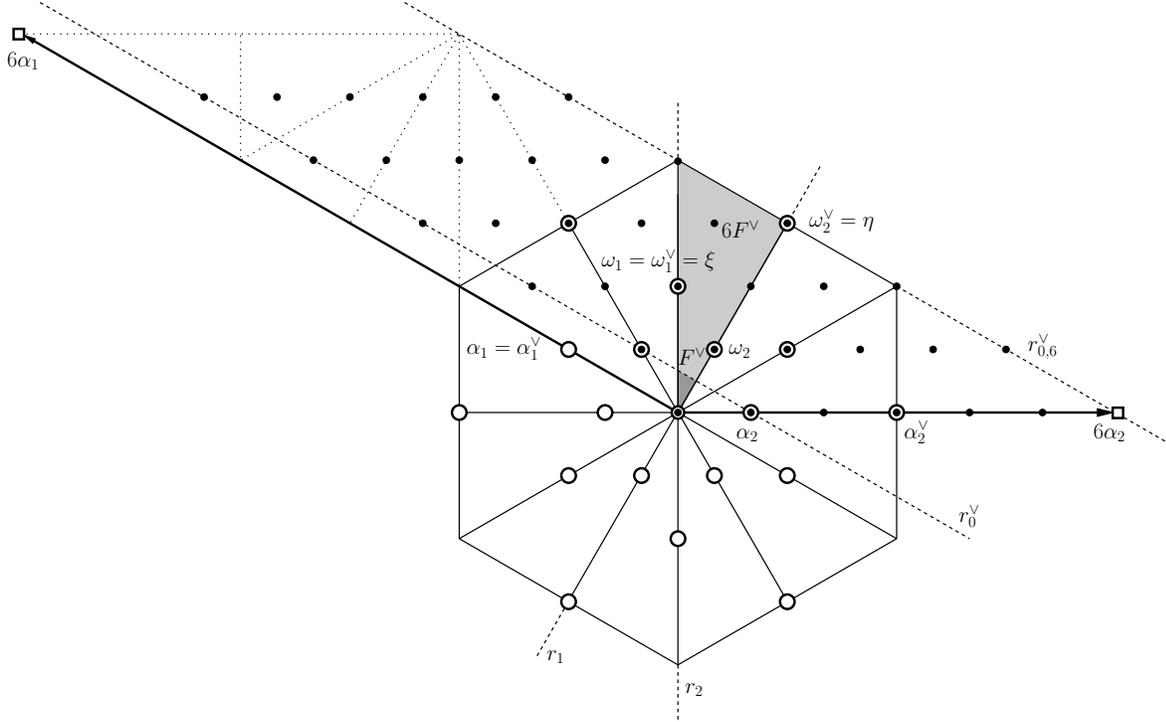}
\caption{The magnified fundamental domain $6F^\vee$ and its boundaries for $G_2$. The darker grey triangle is the fundamental domain $F^{\vee}$ and the lighter grey triangle is the domain $6F^{\vee}$. The coset representatives of $P/6Q$  are shown as $36$ black dots. The seven black points in $6F^\vee$ form the set $\Lambda_6$. The dashed lines represent ``mirrors'' for reflections $r^\vee_0,r^\vee_1$ and $r^\vee_2$ and the affine reflection $r^\vee_{0,6}$.}\label{figG2d}
\end{figure}
\section{Modified multiplication of discretized Weyl-orbit functions}

For any $a\in P_{\R}$, write
\ben  \Phi_\lambda(a)\, \Phi_\mu(a)\ =\ \sum_{\nu\in P_+}\, \langle C \vert CC \rangle _{\lambda,\mu}^\nu\ \Phi_\nu(a)\ ,\label{CCC}
\een
\ben  \Phi_\lambda(a)\, \varphi_{\tilde\mu}(a)\ =\ \sum_{\tilde\nu\in P_{++}}\, \langle S \vert  CS \rangle _{\lambda,\tilde\mu}^{\tilde\nu}\ \varphi_{\tilde \nu}(a)\ ,\label{CSS}
\een
\ben  \varphi_{\tilde \lambda}(a)\, \varphi_{\tilde \mu}(a)\ =\ \sum_{\nu\in P_+}\, \langle C \vert SS \rangle _{\tilde\lambda,\tilde\mu}^\nu\ \Phi_\nu(a)\ , \label{SSC}
\een
for all $\lambda, \mu\in P_+$, and all $\tilde\lambda, \tilde\mu\in P_{++}$.

Similarly, if $\lambda, \mu\in P^M_+$, and all $\tilde\lambda, \tilde\mu\in P^M_{++}$, then
\ben \Phi_\lambda(a)\, \Phi_\mu(a) = \sum_{\nu\in P^M_+}\, {}\sub{M}\langle C \vert CC \rangle _{\lambda,\mu}^{\nu}\ \Phi_\nu(a) , \label{tCCC}
\een
\ben \Phi_\lambda(\tilde a)\, \varphi_{\tilde\mu}(\tilde a)\ =\ \sum_{\tilde\nu\in P^M_{++}}\, {}\sub{M}\langle S \vert CS \rangle _{\lambda,\tilde\mu}^{\tilde\nu}\  \varphi_{\tilde \nu}(\tilde a), \label{tCSS}
\een
\ben  \varphi_{\tilde \lambda}(\tilde a)\, \varphi_{\tilde \mu}(\tilde a)\ =\ \sum_{\nu\in P^M_+}\, {}\sub{M}\langle C \vert  SS \rangle _{\tilde\lambda,\tilde\mu}^{\nu}\  \Phi_\nu(\tilde a) ,\label{tSSC}
\een for any $a\in F_M$ and $\tilde a\in \widetilde F_M$.

Now compare  (\ref{CCC}) and (\ref{tCCC}), for example. The first equation applies for all $\lambda, \mu\in P_+$, and so for the restricted case of $\lambda, \mu\in P^M_+$, as in (\ref{tCCC}). A relation between the decomposition coefficients can be obtained by using the symmetry (\ref{CSWaffhWaff}) on those $\nu\in P_+\backslash P_+^M$ for which $\langle C \vert CC \rangle _{\lambda,\mu}^\nu \not =0$. Weyl-transforming those weights into $P_+^M$ yields
\ben {}\sub{M}\langle C \vert  CC \rangle _{\lambda,\mu}^{\nu}\ =\ \sum_{w\in \widehat W^{\rm aff}_M}\, \langle C \vert CC \rangle _{\lambda,\mu}^{w\nu}\ .\label{CCCw}\een
In similar fashion, \ben {}\sub{M}\langle S \vert  CS \rangle _{\lambda,\tilde\mu}^{\tilde\nu}\ =\ \sum_{w\in \widehat W^{\rm aff}_M}\, \det \widehat\psi (w)\, \langle S \vert  CS \rangle _{\lambda,\tilde\mu}^{w\tilde\nu}\ \label{CSSw}\een
and
\ben {}\sub{M}\langle C \vert  SS  \rangle _{\tilde\lambda,\tilde\mu}^{\nu}\ =\ \sum_{w\in \widehat W^{\rm aff}_M}\, \langle C \vert SS \rangle _{\tilde\lambda,\tilde\mu}^{w\nu}\ \label{SSCw}\een can be derived. These last 3 relations summarize the modified multiplication of discretized orbit functions.  They follow from the affine Weyl symmetries (\ref{CSWaffhWaff}).

\begin{figure}\hspace{1cm}
\resizebox{3.5cm}{!}{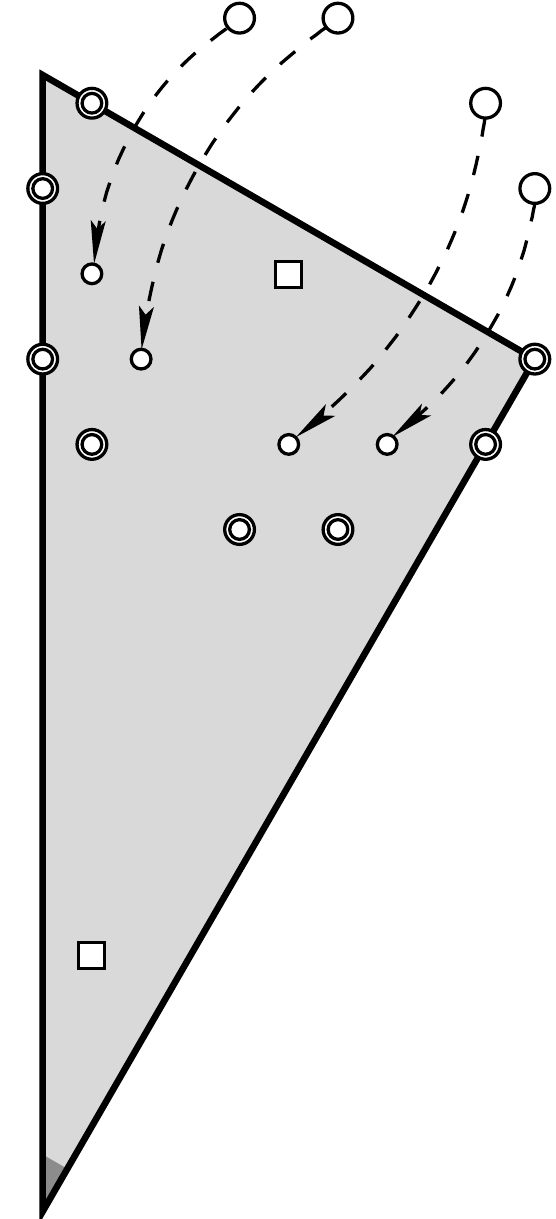}
\hspace{1.8cm}
\resizebox{3.5cm}{!}{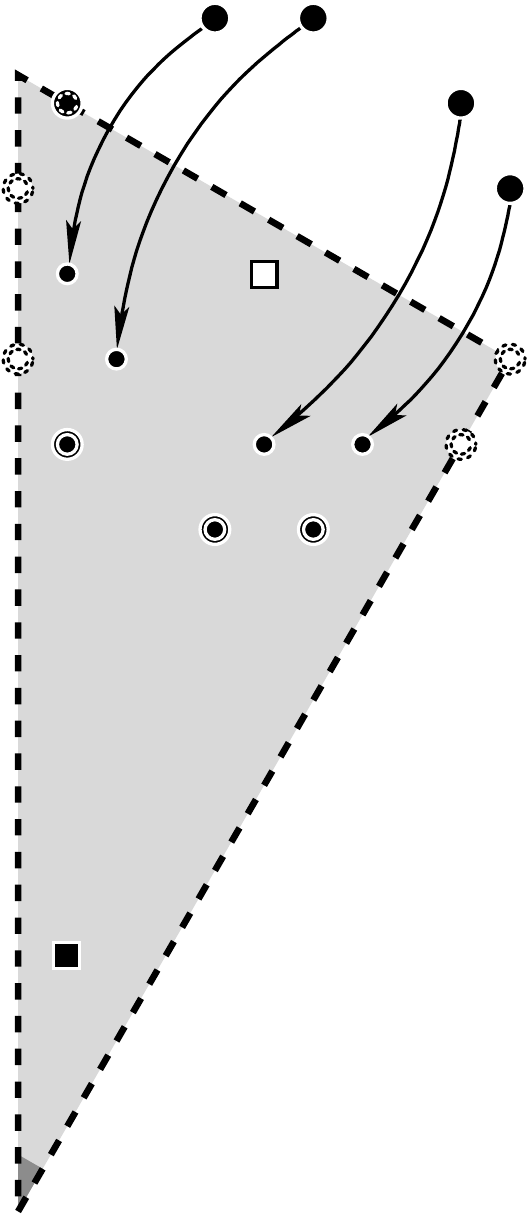}
\hspace{1.8cm}
\resizebox{3.45cm}{!}{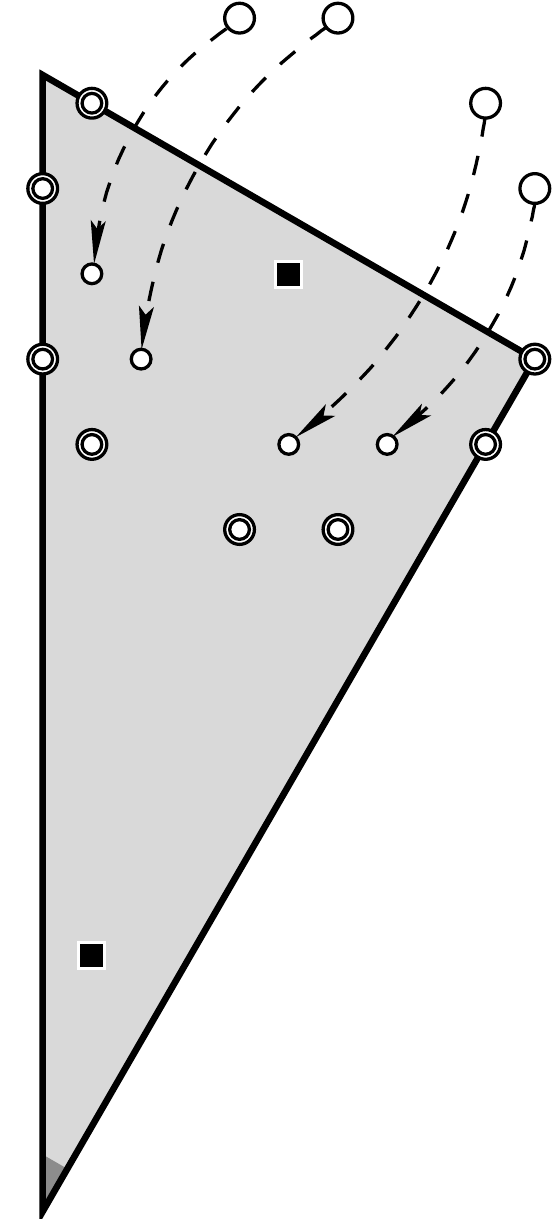}
\caption{The modified multiplication.  The $G_2$ examples of modifications recorded in the text are illustrated here: the product decomposition  (\ref{CCCG}) modified to (\ref{tCCCG}),  (\ref{CSSG}) to (\ref{tCSSG}), and (\ref{SSCG}) to (\ref{tSSCG}), respectively.  The magnified fundamental domain $20F^\vee$ is shaded lightly and the fundamental domain $F^\vee$ more darkly.  The boundaries of $20F^\vee$ are shown as solid or dashed lines, specifying symmetry or antisymmetry of the product terms under reflection across them. Squares indicate the weights labelling the product orbit functions, $(3,5)$ and $(1,1)$, white for Weyl orbit sums ($C$-functions), and black for alternating sums ($S$-functions). Larger circles are drawn at the locations of weights in the Weyl orbit of $(1,1)$, shifted by $(3,5)$. Signed arrows illustrate the modifications, and the smaller circles, the final decompositions in the discretized case. }\label{modmulG2}
\end{figure}

Consider the algebra $G_2$ and two of its weights $(3,5)$ and $(1,1)$. The product decomposition rule
(\ref{CCC}) is of the form
\begin{eqnarray}
\Phi_{(3,5)}\Phi_{(1,1)} =& \Phi_{(4,6)}+\Phi_{(2,9)}+\Phi_{(5,4)}+\Phi_{(6,1)}+\Phi_{(1,10)}+\Phi_{(6,0)} \nn
& +\Phi_{(2,4)}+\Phi_{(4,1)}+\Phi_{(1,6)}+\Phi_{(0,9)}+\Phi_{(5,0)}+\Phi_{(0,10)}
\label{CCCG}
\end{eqnarray}
and its discretized version (\ref{tCCC}) for $M=20$ is of the form
\begin{eqnarray}
\Phi_{(3,5)}{\Phi_{(1,1)}}_{\big|{F_{20}}} =&\Phi_{(4,2)}+\Phi_{(2,5)}+\Phi_{(5,1)}+\Phi_{(6,1)}+\Phi_{(1,7)}+\Phi_{(6,0)}\nn  &+\Phi_{(2,4)}+\Phi_{(4,1)}+\Phi_{(1,6)}+\Phi_{(0,9)}+\Phi_{(5,0)}+\Phi_{(0,10)}.
\label{tCCCG}\end{eqnarray}
The product decomposition rule
(\ref{CSS}) is of the form
\begin{eqnarray}
\Phi_{(3,5)}\varphi_{(1,1)} =& \varphi_{(4,6)}-\varphi_{(2,9)}-\varphi_{(5,4)}+\varphi_{(6,1)}+\varphi_{(1,10)} \nn
& +\varphi_{(2,4)}-\varphi_{(4,1)}-\varphi_{(1,6)}
\label{CSSG}\end{eqnarray}
and its discretized version (\ref{tCSS}) for $M=20$ is of the form
\begin{eqnarray}
\Phi_{(3,5)}{\varphi_{(1,1)}}_{\big|{F_{20}}} =&-\varphi_{(4,2)}+\varphi_{(2,5)}+\varphi_{(5,1)}-\varphi_{(1,7)}\nn &+\varphi_{(2,4)}-\varphi_{(4,1)}-\varphi_{(1,6)}.
\label{tCSSG}\end{eqnarray}
 The product decomposition rule
(\ref{SSC}) is of the form
\begin{eqnarray}
\varphi_{(3,5)}\varphi_{(1,1)} =& \Phi_{(4,6)}-\Phi_{(2,9)}-\Phi_{(5,4)}+\Phi_{(6,1)}+\Phi_{(1,10)}-\Phi_{(6,0)} \nn
& +\Phi_{(2,4)}-\Phi_{(4,1)}-\Phi_{(1,6)}+\Phi_{(0,9)}+\Phi_{(5,0)}-\Phi_{(0,10)}
\label{SSCG}\end{eqnarray}
and its discretized version (\ref{tSSC}) for $M=20$ is of the form
\begin{eqnarray}
\varphi_{(3,5)}{\varphi_{(1,1)}}_{\big|{F_{20}}} =&\Phi_{(4,2)}-\Phi_{(2,5)}-\Phi_{(5,1)}+\Phi_{(6,1)}+\Phi_{(1,7)}-\Phi_{(6,0)}\nn &+\Phi_{(2,4)}-\Phi_{(4,1)}-\Phi_{(1,6)}+\Phi_{(0,9)}+\Phi_{(5,0)}-\Phi_{(0,10)}.
\label{tSSCG}\end{eqnarray}
Diagrams depicting the transition from ordinary to modified multiplication are contained in Figure \ref{modmulG2}.

\section{Galois symmetry of Weyl-orbit functions}

Let $N$ denote the minimum positive integer such that \bens   \left( e^{2\pi i\langle \lambda, a\rangle} \right)^{N}\ =\ e^{2\pi i\langle {N}\lambda, a\rangle}\ =\ 1\ , %\label{Ndef}
\eens for all $\lambda\in \Lambda_M,\ a\in F_M$.

Suppose $\gcd(\ell, N)=1$ for $\ell\in\N$.  Define the Galois transformation $t_\ell$ by
\bens t_\ell\left( e^{2\pi i\langle \lambda, a \rangle}  \right) :=\ e^{2\pi i \ell \langle \lambda, a \rangle}\ \ ,%\label{tell}
\eens and extend it linearly to sums of such terms.  This transformation swaps one primitive root of unity for another.

Applying the Galois transformation to the orbit functions, one gets \bea     t_\ell\bigg( \Phi_\lambda(a) \bigg)  = \Phi_{\ell\lambda}(a)\ = \Phi_\lambda(\ell a)\ , \nn t_\ell\bigg( \varphi_{\tilde\lambda}(\tilde a) \bigg)  = \varphi_{\ell\tilde\lambda}(\tilde a) = \varphi_{\tilde\lambda}(\ell \tilde a)\ . \label{CSgal}\eea Now suppose that $\lambda\in P_+^M$, $\tilde\lambda\in P^M_{++}$, $a\in F_M$, and $\tilde a\in \widetilde F_M$.  Multiples of these weights by a factor of $\ell$ will not also, in general, be part of the same sets. They can all, however, be moved there by appropriate elements of the relevant affine Weyl group: \bea  \hat w_\ell[\lambda]\, \left(\, \ell\, \lambda \,\right)\  =: t_\ell[\lambda]\ \in\ P^M_+\  \ ,\qquad \hat w_\ell[\lambda]\, \in\ \widehat W^{\rm aff}_M ,\nn
\hat w_\ell[\tilde\lambda]\, \big(\, \ell\, \tilde\lambda \,\big)\  =: t_\ell[\tilde\lambda]\  \in\  P^M_{++}\   ,\qquad \hat w_\ell[\tilde\lambda]\, \in\ \widehat W^{\rm aff}_M,\nn w_\ell[a]\, \left(\, \ell\, a \,\right)\  =: t_\ell[a]\ \in\ F_M\ \  ,\qquad w_\ell[a]\, \in\ W^{\rm aff}, \label{well}\\
w_\ell[\tilde a]\, \left(\, \ell\, \tilde a \,\right)\  =: t_\ell[\tilde a]\ \in\ \widetilde F_M\ \  ,\qquad w_\ell[\tilde a]\, \in\ W^{\rm aff} .\nonumber\eea Combining the affine Weyl symmetries (\ref{CSWaffhWaff}) with (\ref{CSgal}), (\ref{well}) then gives the Galois symmetry of the orbit functions: \bea  t_\ell\bigg( \Phi_\lambda(a) \bigg)  =  \Phi_{t_\ell[\lambda]}(a) = \Phi_\lambda(t_\ell[a])\ , \nn t_\ell\bigg( \varphi_{\tilde\lambda}(\tilde a) \bigg) \ =\ \hat\epsilon_\ell[\tilde\lambda]\, \varphi_{t_\ell[\tilde\lambda]}(\tilde a)\ =\ \epsilon_\ell[\tilde a]\, \varphi_{\tilde\lambda}(t_\ell[\tilde a])\ .     \label{tlCS}\eea
Here we have defined the signs \beas \hat\epsilon_\ell[\tilde\lambda]\ :=\ \det\widehat \psi (\hat w_\ell[\tilde\lambda]) \ ,\ \ \hat w_\ell[\tilde\lambda]\, \in\ \widehat W^{\rm aff}_M ,\nn    \epsilon_\ell[\tilde a]\ :=\ \det\psi\big(\, w_\ell[\tilde a]  \,\big)\ ,\ \ w_\ell[\tilde a]\, \in\ W^{\rm aff}.%\label{signs}
\eeas
The Galois symmetries are interesting relations among the values of discretized Weyl-orbit functions. Each $\ell$ coprime to $N$ yields a permutation of the weights relevant to the orbit functions, both the argument weights and the label weights, and each such Galois transformation produces relations (\ref{tlCS}).

Galois symmetry also produces relations involving the decomposition coefficients discussed above.  For example, because the Galois transformation exchanges one root of unity for another, and because the coefficients ${}\sub{M}\langle C \vert\  SS \rangle_{\tilde\lambda,\tilde\mu}^{\nu}$ are rational, equations (\ref{tSSC}) and (\ref{tlCS}) imply \bens  t_\ell\bigg(\, \varphi_{\tilde \lambda}(\tilde a)\,\bigg)\, t_\ell\bigg(\,\varphi_{\tilde \mu}(\tilde a)\,\bigg) = \sum_{\nu\in P^M_+}\, {}\sub{M}\langle C \vert SS \rangle_{\tilde\lambda,\tilde\mu}^{\nu}\  t_\ell\bigg(\, \Phi_\nu(\tilde a) \,\bigg)\ .%\label{tlSSC}
\eens Applying again (\ref{tlCS}) then yields \bens  \hat\epsilon_\ell[\tilde\lambda]\, \varphi_{t_\ell[\tilde \lambda]}(\tilde a)\, \hat\epsilon_\ell[\tilde\mu]\, \varphi_{t_\ell[\tilde \mu]}(\tilde a)  = \sum_{\nu\in P^M_+}\, {}\sub{M}\langle C \vert SS \rangle_{\tilde\lambda,\tilde\mu}^{\nu}\  \Phi_{t_\ell[\nu]}(\tilde a) \ ,%\label{telltSSC}
\eens
so that
\bea  \hat\epsilon_\ell[\tilde\lambda]\, \hat\epsilon_\ell[\tilde\mu]\, \sum_{\nu\in P^M_+}\, {}\sub{M}\langle C \vert SS \rangle_{t_\ell[\tilde\lambda],t_\ell[\tilde\mu]}^{\nu}\  \Phi_\nu(\tilde a)  \, \nn \QQ =\, \sum_{\nu\in P^M_+}\, {}\sub{M}\langle C \vert SS \rangle_{\tilde\lambda,\tilde\mu}^{\nu}\, \, \Phi_{t_\ell[\nu]}(\tilde a) \ .\label{SSCtellSSC}\eea
Since both sides of equation (\ref{SSCtellSSC}) are zero for $a\in F_M \setminus \widetilde F_M$, it is valid for all $a\in F_M$.
Discrete orthogonality relations for $\Phi_\nu$ over $F_M$ from \cite{disc} then produce \ben\hat\epsilon_\ell[\tilde\lambda]\, \hat\epsilon_\ell[\tilde\mu]\, {}\sub{M}\langle C \vert SS \rangle_{t_\ell[\tilde\lambda],t_\ell[\tilde\mu]}^{t_\ell[\nu]} = {}\sub{M}\langle C \vert SS \rangle_{\tilde\lambda,\tilde\mu}^{\nu}\,\, \ .\label{NSSCtellSSC}
\een
Similar relations hold for the other decomposition coefficients of (\ref{tCCC}-\ref{tSSC}).
\begin{figure}\hspace{1cm}
\resizebox{3.55cm}{!}{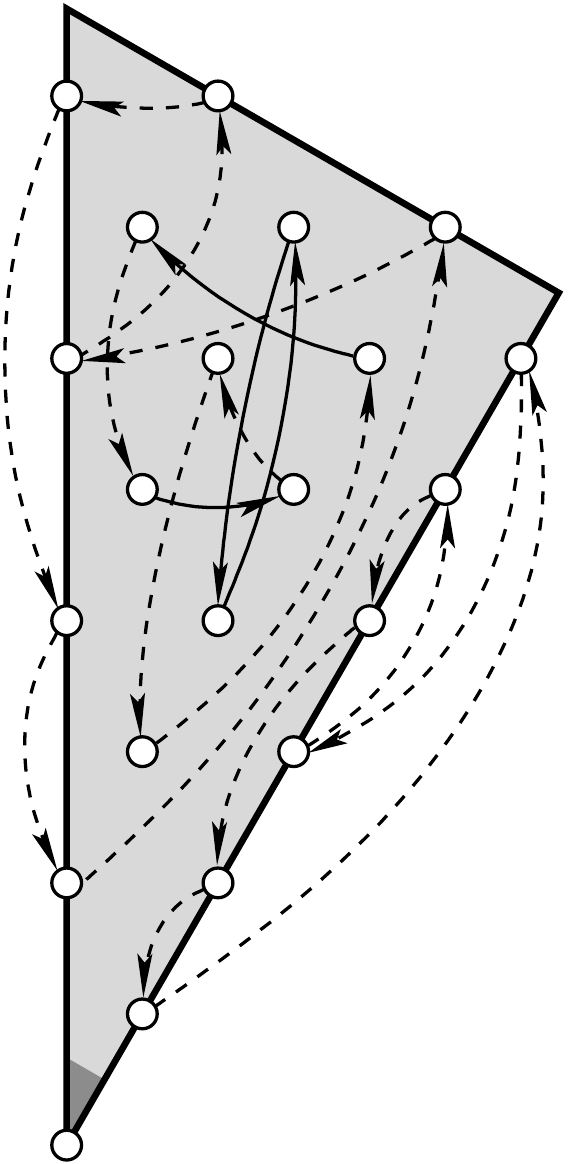}
\hspace{1.8cm}
\resizebox{3.45cm}{!}{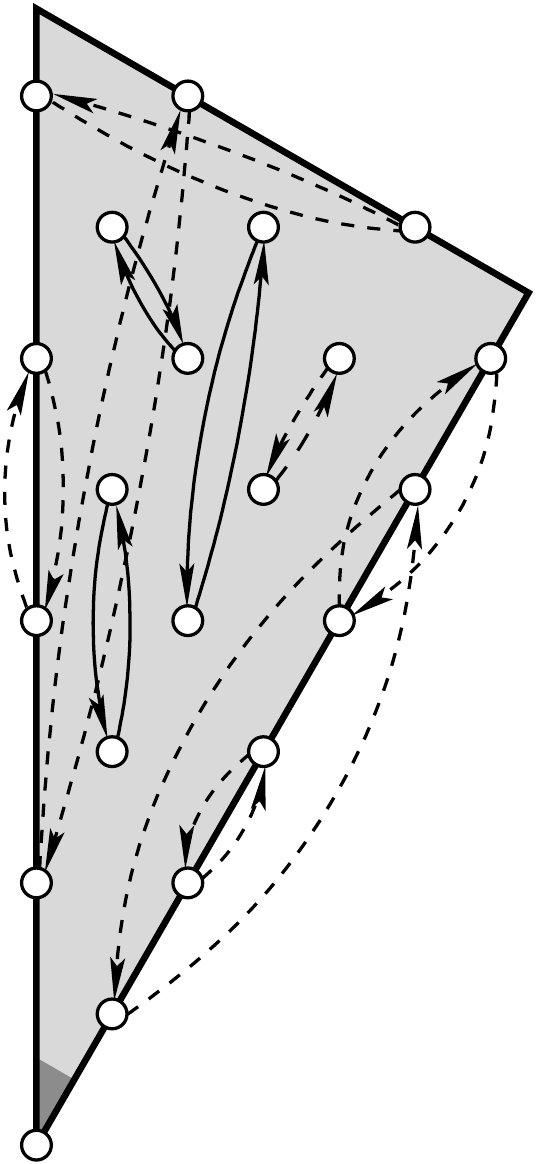}
\hspace{1.8cm}
\resizebox{3.5cm}{!}{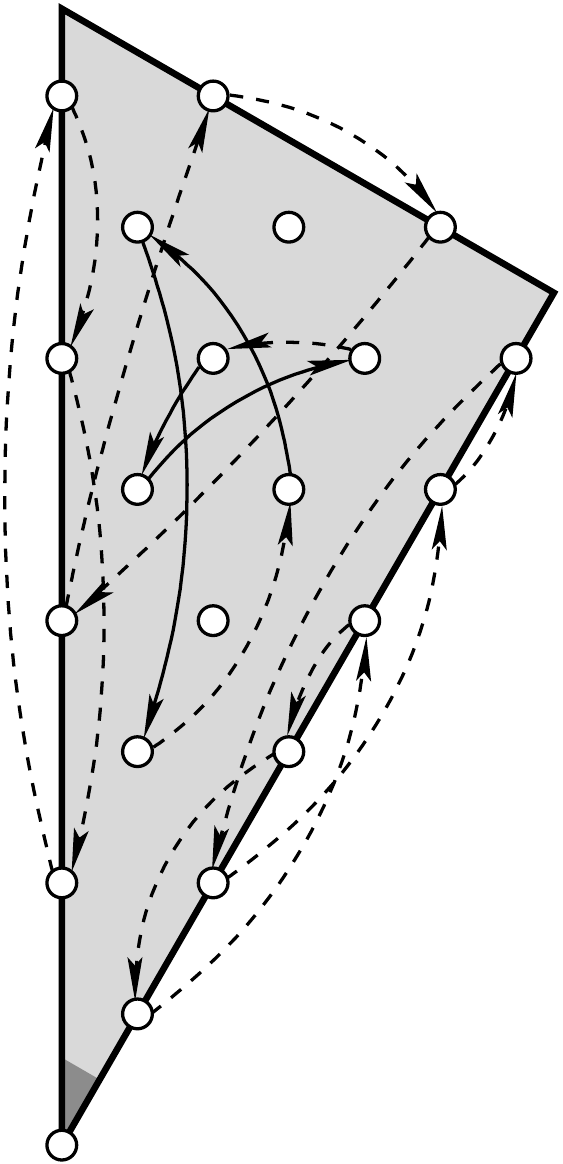}
\caption{Galois transformations and signs. Pictured are the transformations $t_\ell[\lambda]$ and the signs $\hat\epsilon_\ell[\lambda]$ of (\ref{tlCS}) for $G_2$ and $\lambda\in\Lambda_M$, with $M=13$, and $\ell=7,8,9$, respectively. The scaled-up fundamental domain $13F^\vee$ is shaded lightly and the fundamental domain $F^\vee$ more darkly. The white circles indicate weights in  $\Lambda_{13}$.  Arrows connect $\lambda\in \Lambda_{13}$ to $t_\ell[\lambda]$. Solid arrows correspond to $\hat\epsilon_\ell[\lambda]=+1$, while dashed arrows in the interior of $\Lambda_{13}$ signify $\hat\epsilon_\ell[\lambda]=-1$. Dashed arrows are also used on the boundary, where $\hat\epsilon_\ell[\lambda]$ is undefined. }\label{GalG2}
\end{figure}

\section{Wess-Zumino-Novikov-Witten conformal field theories: affine modular data}

Modified multiplication and Galois symmetries appear in the study of modular data for rational conformal field theories. In particular, objects related to the modular data of Wess-Zumino-Novikov-Witten (WZNW) models bear a striking resemblance to the discretized Weyl-orbit functions discussed here.  This modular data is associated with the affine Kac-Moody algebras of untwisted type, at a fixed level, and so is also called affine modular data.

The relation between affine modular data and Weyl-orbit functions, of both $S$ and $C$ type, was already exploited in \cite{GJW}.

As for any rational conformal field theory, the WZNW 1-loop partition function is invariant under the modular group $SL(2;\Z)$, with generators $S$ and  $T$. This is possible because the 1-loop conformal blocks can be identified with the characters of the untwisted affine Kac-Moody algebra that is the algebra of the loop group of $G$, at a fixed {\it level}, usually denoted $k$. The relevant affine Kac-Moody algebra at fixed level $k\in\N_0$ has $g_n$ as its horizontal subalgebra, and is often denoted $g_{n,k}$. The affine characters form a finite-dimensional representation
of the modular group \cite{KaP}. The generators $S$ and $T$ of the modular group can therefore be represented by finite-dimensional matrices.

The matrix $S$ involves the antisymmetric orbit functions discussed above.  To write its formula, we need to define a few more Lie-algebraic quantities. If the highest root of $g_n$ obeys $\xi\ =\ \sum_{j\in I} m'_j\,\alpha_j^\vee$, then the $m'_j$ are known as co-marks. Their sum defines the dual Coxeter number $h^\vee := 1 + \sum_{j\in I} m'_j$. Let $\Delta_+$ denote the set of positive roots of $g_n$ and put $M' := k+h^\vee$.  The affine characters can be labeled by weights in $P^{M'}_{++}$, and the Kac-Peterson modular $S$ matrix has the form
\ben  S_{\lambda,\mu} =  i^{\Vert \Delta_+ \Vert}\, \vert P/Q^\vee\vert^{-\frac 1 2}\, (M')^{-\frac r 2} \,\,  \sum_{w\in W}\, (\det w)\, e^{-2\pi i \langle w\lambda, \mu\rangle/{M'}}\ \ , \label{KPmatrixS}\een  for $\lambda,\mu\in P^{M'}_{++}$.
We can therefore rewrite the Kac-Peterson formula as \bea  S_{\lambda,\mu} =  R_{M'}\,  \sum_{w\in W}\, (\det w)\, e^{-2\pi i \langle w\lambda, \mu/{M'} \rangle}\ =\ R_{M'}\, \varphi_\lambda(-\mu')\ , \nn  \qq {\rm with}\ \ \mu'=\mu/M'\in P_{++}^{M'}/M'\ \  {\rm and}\ \  \lambda\in P_{++}^{M'}\ .\label{KPorbitS}\eea $R_{M'}$ is the constant written out in (\ref{KPmatrixS}).

Notice that even if we identify $M'$ and $M$, the weights involved in the $S$-functions of (\ref{KPorbitS}) are different from those implicated above.  Normally, for the orbit functions $\varphi_\lambda(\mu)$ we have $\lambda\in \widetilde\Lambda_M = P_{++}^M$ and $\mu\in \widetilde F_M = P^{\vee\, M}_{++}/M$. The symmetry of the matrix with entries (\ref{KPmatrixS}) is consistent with this difference. Also consistent is the affine Weyl group symmetry \cite{KaP, K} of the modular matrix $S$:
\bens  S_{w\lambda, \mu} = S_{\lambda, w\mu} = \det \psi (w)\cdot S_{\lambda,\mu}\ , \ \ \forall\, w\in W^{\rm aff}_{M'}\ . %\label{WaffSmod}
\eens

The affine Weyl symmetry still gives rise to a Galois symmetry \cite{CG}, an important property of rational conformal field theories.  The Galois symmetry of WZNW models is the motivation for the Galois symmetry discussed above.

The modular matrix $S$ is vital in rational conformal field theories, not least because it enters the Verlinde formula for fusion coefficients.  For the WZNW models, the fusion coefficients are expressed in terms of the Kac-Peterson modular matrix $S$:  \bens {}^{(k)}N_{\lambda, \mu}^\nu\ =\ \sum_{\sigma\in P_{++}^{M'}}\, \frac{ S_{\lambda, \sigma}\, S_{\mu, \sigma}\, S_{\nu, \sigma}^*}{S_{\rho, \sigma}}\ . %\label{VmodS}
\eens Here $*$ denotes complex conjugation, $\rho = \sum_{j\in I} \omega_j$ is the Weyl vector, and $M'=k+h^\vee$, as above. The modular matrix $S$ is unitary, as well as symmetric, and so the Verlinde formula can be re-expressed as
\ben \left(\frac{ S_{\lambda, \sigma} }{S_{\rho, \sigma}}\right)\, \left(\frac{ S_{\mu, \sigma} }{S_{\rho, \sigma}}\right) =  \sum_{\nu\in P_{++}^{k+h^\vee}}\, {}^{(k)}N_{\lambda, \mu}^\nu\, \left(\frac{ S_{\nu, \sigma} }{S_{\rho, \sigma}}\right) \ . \label{Vchi}
\een

The ratios of the previous equation, are ratios of antisymmetric Weyl-orbit sums, and so are (discretized) Weyl characters of integrable, highest-weight representations of $g_n$: \bens  \left(\frac{ S_{\lambda, \sigma} }{S_{\rho, \sigma}}\right) = \chi_\lambda(\sigma)\ .%\label{KPratio}
\eens Here $\chi_\lambda$ denotes the character of the representation with highest weight $\lambda-\rho\in P_+^k$.  Now we can indicate the WZNW motivation for the modified multiplication discussed above.

Since products of characters decompose as tensor products of representations do, we can write \ben  \chi_\lambda(\sigma)\, \chi_\mu(\sigma) = \sum_{\phi\in P_{++}}\, T_{\lambda, \mu}^\phi\, \chi_\phi(\sigma)\ ,   \label{chiT}\een where $T_{\lambda, \mu}^\phi$ is the tensor product coefficient.  But (\ref{Vchi}) is just \ben  \chi_\lambda(\sigma)\, \chi_\mu(\sigma) = \sum_{\nu\in P_{++}^{k+h^\vee}}\, {}^{(k)}N_{\lambda, \mu}^\nu\, \chi_\nu(\sigma)\ .   \label{chiN}\een  Using the affine Weyl symmetry to compare (\ref{chiN}) and (\ref{chiT}), we find \cite{K, W} \bens {}^{(k)}N_{\lambda, \mu}^\nu = \sum_{w\in W^{\rm aff}_{M'}}\, \det\psi ( w)\cdot  T_{\lambda, \mu}^{w\nu}\ .  % \label{NdetwT}
\eens  While (\ref{chiT}) is valid for any weight $\sigma\in P_\R$, (\ref{chiN}) is true only for the discretized weights $\sigma\in P_{++}^{k+h^\vee}$.  The discretization results in a modified multiplication for the characters, so that the tensor product coefficients are replaced by the fusion coefficients. The modified multiplications of discretized Weyl-orbit functions work essentially the same way.

\section{Conclusion}

Our main result is the recognition of the resemblance between affine (WZNW) modular data, and discretized Weyl-orbit functions. We would like to mention here that the modular data and discretized orbit functions have already appeared together in \cite{GJW}, which was part of the motivation for this paper. In that work, discretized $C$-functions, in conjunction with the modular $S$-matrix, gave rise to a simple formula for the weight multiplicities of any simple Lie algebra.

Here we worked out 2 consequences for discretized orbit functions, by imitating the derivation of 2 well-known, important properties of the affine modular data.  First, just as affine (WZNW) fusion is a truncated version of the tensor product of Lie algebra representations \cite{K,W}, so is the product of continuous-valued orbit functions modified when the discretized orbit functions are used.  Equations (\ref{CCCw}-\ref{SSCw}) give the precise relations between the corresponding decomposition coefficients. Second, the analogue for the discretized orbit functions of the Galois symmetry of the affine modular $S$-matrix \cite{CG} was displayed, in equations (\ref{NSSCtellSSC}).

There should be other knowledge of affine modular data that can teach us about discretized orbit functions. For example, the Galois symmetry of the orbit functions may be extendable to a quasi-Galois symmetry, since the affine modular data has that property \cite{quasiG}.\footnote{MW thanks J\"urgen Fuchs for pointing out this possibility.} The concept of fusion generator and the phenomenon of fixed-point factorization found in \cite{GW} may also transfer to the discretized orbit functions.

In the opposite direction, we also hope that the orbit functions can help us learn more about affine modular data, and the relevant WZNW conformal field theories. Furthermore, a standard motivation for studies of WZNW models is that some results may possibly be generalized to larger classes of rational conformal field theories. All that, however, is beyond the scope of this paper.

%\begin{acknowledgments}
\vskip1cm
\noindent{\bf Acknowledgments}\hfb
This research was supported in part by a Discovery Grant (MW) from the Natural Sciences and Engineering Research Council (NSERC) of Canada. JH gratefully acknowledges the support of this work by RVO68407700.
%\end{acknowledgments}

%\bibliography{polytopebib}

\vskip1cm
\noindent\today
\end{document}